# An Energy-Efficient Generic Accuracy Configurable Multiplier Based on Block-Level Voltage Overscaling

Ali Akbar Bahoo, Omid Akbari, Muhammad Shafique

**Abstract**—Voltage Overscaling (VOS) is one of the well-known techniques to increase the energy efficiency of arithmetic units. Also, it can provide significant lifetime improvements, while still meeting the accuracy requirements of inherently error-resilient applications. This paper proposes a generic accuracy-configurable multiplier that employs the VOS at a coarse-grained level (block-level) to reduce the control logic required for applying VOS and its associated overheads, thus enabling a high degree of trade-off between energy consumption and output quality. The proposed configurable Block-Level VOS-based (BL-VOS) multiplier relies on employing VOS in a multiplier composed of smaller blocks, where applying VOS in different blocks results in structures with various output accuracy levels. To evaluate the proposed concept, we implement 8-bit and 16-bit BL-VOS multipliers with various blocks width in a 15-nm FinFET technology. The results show that the proposed multiplier achieves up to 15% lower energy consumption and up to 21% higher output accuracy compared to the state-of-the-art VOS-based multipliers. Also, the effects of Process Variation (PV) and Bias Temperature Instability (BTI) induced delay on the proposed multiplier are investigated. Finally, the effectiveness of the proposed multiplier is studied for two different image processing applications, in terms of quality and energy efficiency.

——————————— ◆ ———————————

## 1. Introduction

TODAY, multipliers, being one of the basic building blocks of digital systems, play an important role in determining the required power/energy budget for different applications, such as machine learning, digital signal processing (DSP), and multi-media processing applications [1]. This is more critical for the applications executing on portable digital devices, where the power/energy constraint is more stringent [2]. Therefore, designing an energy-efficient multiplier with high computational efficiency essentially improves the energy efficiency of digital systems employed in a wide range of applications. In general, the main criteria for designing such a multiplier include reducing power consumption and increasing performance, as well as offering lower area and increased lifetime. However, with the technology scaling and reducing sizes of transistors, reducing aging and improving reliability have also become vital issues for designers [3]. Reducing the power consumption of circuits leads to their longer lifetime and decelerates the aging phenomenon, resulting in higher reliability [2]. Dynamic Voltage and Frequency Scaling (DVFS) is one of the most widely-used techniques to reduce both dynamic and static power based on the application requirements. The Voltage Overscaling (VOS) stretches the concept of DVFS to aggressively scale down the voltage for ultra-low power consumption, but may lead to timing violations that may not be acceptable for certain applications [4].

However, several applications, from domains like digital signal processing, audio and video processing, computer vision, machine learning, and data mining, are inherently error-resilient and can still produce useful results even in the presence of moderate errors. This concept has been leveraged by the emerging trend of Approximate Computing (AxC) for designing energy-efficient systems while exploiting the inherent error-resiliency of applications [2]. The goal is to have maximal energy, latency and/or area savings under a tolerable error constraint considering the Quality-of-Service (QoS) requirements of the given application.

For AxC, VOS has been employed in different levels of hardware design, from the circuit to the system level, e.g., [3], [4], [5], [6], and [7]. The state-of-the-art works typically employ VOS at a fine-grained level [3], which limits the benefits of AxC and imposes significant overheads due to many different control signals that determine the voltage level of multiplier components (see Fig. 1.a). In this paper, we propose a generic Block-Level VOS (BL-VOS)-based accuracy configurable multiplier to achieve energy-efficient approximate multipliers with improved lifetime, as well as lower area and design complexity. To achieve BL-VOS multiplier, we apply the VOS technique to a multiplier composed of smaller blocks [1] (see Fig. 1.b). The bit-width of these blocks along with the voltage level of each block are the knobs that provide an accuracy configurable multiplier. These knobs are used to create an appropriate tradeoff between accuracy and energy efficiency while offering a multiplier with an improved lifetime and mitigate-

- A. Bahoo is with the Department of Electrical and Computer Engineering, University of Eyvanekey, Semnan 1146-35915, Iran (e-mail: ali.bahoo@eyc.ac.ir).
- O. Akbari is with the Department of Electrical and Computer Engineering, Tarbiat Modares University, Tehran 14115-111, Iran (e-mail: o.akbari@modares.ac.ir).
- M. Shafique is with the Division of Engineering, New York University Abu Dhabi (NYU AD), Abu Dhabi 129188, United Arab Emirates (e mail: muhammad.shafique@nyu.edu)

"This work was supported in parts by the NYUAD's Research Enhancement Fund (REF) Award on "eDLAuto: An Automated Framework for Energy-Efficient Embedded Deep Learning in Autonomous Systems", and by the NYUAD Center for CyberSecurity (CCS), funded by Tamkeen under the NYUAD Research Institute Award G1104."

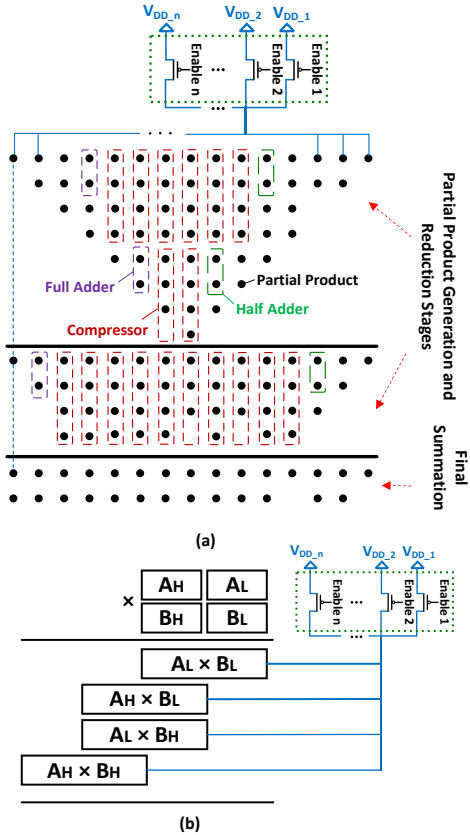

Fig. 1. (a) Applying the VOS at a fine level to the different components of a Dadda multiplier [3], (b) applying the VOS at a coarse level to the different blocks of a multiplier built out of smaller blocks.

ed aging.

**The contributions of this paper are as follows:**

1) We propose a systematic method for applying VOS to the block-based multipliers to achieve a trade-off between output quality and energy efficiency, as well as for improving the lifetime.
2) We propose an accuracy configurable multiplier with the ability to adjust the accuracy during runtime.
3) Leveraging VOS at block-level to achieve a lower overhead multiplier compared to the state-of-the-art.
4) Analysis of the proposed generic accuracy configurable multiplier for different voltage levels and bit widths of blocks for the aforementioned trade-off.
5) Studying the effects of process variations on the accuracy of the proposed approximate multiplier.
6) Investigating the impact of the VOS on the lifetime of the proposed accuracy configurable multiplier.

*Paper Organization*: In Section 2, prior works related to approximate computing and VOS-based arithmetic units are reviewed. The background on lifetime, aging effects, error metrics employed in the AxC literature, and block-based multipliers are presented in Section 3. The structure of the proposed block-level VOS-based generic accuracy configurable multiplier is described in Section 4. The simulation results of different structures of the multipliers are presented in Section 5. Finally, this paper is concluded in Section 6.

## 2. RELATED WORKS

In this section, we review the prior works on approximate computing. Mainly, we focus on approximate multipliers constructed through logic manipulations, and the most relevant researches that employed VOS in the structure of multipliers.

### 2.1. Approximate Computing

Recently, AxC has received great attention and has been employed at different hardware levels, from the circuit [8] to the architecture levels (see [5] and [9]). Comprehensive reviews on the approximate computing paradigm can be found in [10].

In [8], the AxC method was used at the transistor level to design low-power approximate full adders. In [5], a framework was proposed to find the most appropriate micro-architecture of a hardware accelerator for dynamic voltage overscaling (DVS). Also, a controller was used to manage the supply voltage based on the workload. In [9], an approximate coarse-grained reconfigurable architecture (X-CGRA) was proposed. The X-CGRA consists of a RISC processor, an approximate reconfigurable array (X-RA) of Quality-Scalable Processing Elements (QSPEs), and some memory units. Moreover, a mapping method was proposed to map different applications with the desired quality on the X-CGRA. The quality configuration of X-CGRA provides a runtime trade-off between accuracy and energy efficiency.

Some recent works were focused on testing the digital approximate circuits, where a comprehensive survey can be found in [40]. In [41], generating test patterns for digital circuits were discussed. This paper introduces the method of choosing a test subset among the all-possible test sets (test vectors) such that it ensures that no fault will result in an error greater than the given tolerable range.

### 2.2. Logic Manipulation based Approximate Multipliers

In [1], Kulkarni et. al. proposed an approximate 2×2 multiplier based on logic simplifications through Karnaugh-Map. The proposed approximate multiplier takes up half the area compared to the accurate version. This multiplier is then used for constructing larger approximate multipliers, e.g., 8×8 and 16×16 multipliers. In [11], the Broken-Array Multiplier (BAM) was proposed in which some carry-save adder (CSA) cells of the rows and least significant columns of the partial products (PPs) array are removed to reduce the logic complexity at the cost of minor quality loss. These modifications result in area and energy consumption reduction. In [12], a library of approximate adders and multipliers was generated by using Cartesian Genetic Programming (CGP), where the integer chromosomes show the structure of the approximate arithmetic units.

TABLE I  COMPARISON OF THE DIFFERENT STATE-OF-THE-ART APPROXIMATE MULTIPLIERS

| Structure | Technology Node (nm) Used in the Reference | VOS | Quality Configurable | Aging Investigation | Block-Level Configurability |
|---|---|---|---|---|---|
| BAM [11] | CMOS 130 | ✗ | ✗ | ✗ | ✓ |
| EVO [12] | TSMC 180 | ✗ | ✗ | ✗ | ✓ |
| PPAM [13] | TSMC 65 | ✗ | ✗ | ✗ | ✗ |
| YUS [17] | NanGate 45 | ✗ | ✓ | ✗ | ✗ |
| YUS-V2 [18] | NanGate 45 | ✗ | ✓ | ✗ | ✗ |
| X-Dadda [3] | FinFET 15 | ✓ | ✓ | ✓ | ✗ |
| BL-VOS | FinFET 15 | ✓ | ✓ | ✓ | ✓ |

In this method, mutation operation was used to generate new populations of the genetic algorithm. Also, the fitness function is defined based on mean relative error (MRE). In [13], the partial product perforation was employed to construct approximate multipliers.

In [14], some approximate compressors were proposed based on AND-OR gates and employed in the partial product reduction stage of a multiplier. In [15], a static segment method (SSM) was proposed that multiply two m-bit segments instead of multiplying two n-bit input operands, where $n/2 \leq m$ and the value of $m$ is determined by leading one detector. Thus, less energy is consumed compared to an n-bit multiplier.

In [16], two 4:2 approximate compressors were proposed. These compressors were employed at the reduction stage of Dadda multipliers. Four quality configurable 4:2 compressors were proposed in [2]. These compressors provide the ability to switch between exact and approximate modes at run-time. Afterward, these compressors were employed in the partial product reduction stage of Dadda multipliers to build accuracy configurable multipliers. In [17], a configurable approximate multiplier was proposed. In this multiplier, a quality configurable carry-maskable adder (CMA) was used at the final summation stage of the multiplier. Then, by changing the length of the carry propagation in the CMA, the accuracy of the multiplier is varied during run-time. In addition, some approximate compressors are used in the partial product reduction stage of the multiplier. Similarly, an accuracy configurable approximate multiplier composed of approximate compressors and CMA was proposed in [18]. However, in [18], using a simpler approximate tree compressor resulted in a lower area and power consumption compared to [17].

### 2.3. VOS-based Accuracy Configurable Multipliers

In [3], VOS technique was used to build accuracy configurable Dadda multipliers. Also, the four-bit truncation method was employed for further efficiency gains. In this method, VOS was used at a fine-grained level (see Fig. 1.a), where a number of voltage levels were used at the partial product generation and reduction, and final summation stages. Also, level shifters were used to transfer signals from low voltage to high voltage. However, applying the VOS at a fine-grained level may limit the benefit of AxC due to increased complexity of the circuit design. Specifically, as the size of the multiplier is increased (e.g., 16 and 32-bit multipliers), this complexity significantly reduces the AxC advantages. In [6], a low power probabilistic floating-point multiplier was proposed based on biased-VOS (Bi-VOS), where the VOS is applied to the full adders in each column of an array multiplier. This method applied voltage levels to the circuit according to an energy budget constraint. Also, a sleep scheme was employed to achieve further energy efficiency, in which some gates are put in a non-operational state. In [7], based on an analytical method, an energy assignment flow for the probabilistic multipliers was proposed. This flow targets to create a trade-off between the energy consumption and error induced by probabilistic multipliers.

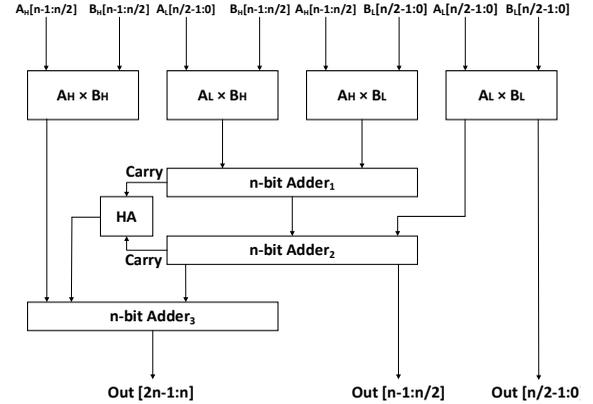

Fig. 2. General structure of a block-based multiplier.

TABLE I provides a summary of the key characteristics of the state-of-the-art multipliers. These multipliers are later used in Subsection 5.4 for benchmarking our proposed design. Unlike prior works that employed the VOS technique at a fine-grained level, we propose a systematic method to employ VOS at a coarse-grained level. In this regard, the resultant multiplier offers higher energy efficiency and lifetime, as well as a simpler design with lower overheads.

## 3. BACKGROUND

In this section, first, the structure of a generic block-based multiplier is discussed. Next, aging factors and their effects on the performance of transistors are presented. Based on the relation between the aging and voltage level of the circuits, it is shown that the lower voltage level results in a higher lifetime.

### 3.1. Block-Based Multiplier

The general structure of a block-based multiplier is depic-ted in Fig. 2. As shown in this figure, the $n$-bit $A$ and $B$ inputs are split into two higher and lower parts, including ($A_H$, $B_H$) and ($A_L$, $B_L$), respectively [1]. Next, these parts are multiplied by each other in four separated blocks, including $A_L \times B_L$, $A_H \times B_L$, $A_L \times B_H$, and $A_H \times B_H$. Finally, three $n$-bit adders and one-half adder (HA) are used to calculate the $2n$-bit output ($Out$). It should be noted that, without loss of generality, each sub-block of this multiplier can be implemented using a different type of multiplier, e.g., Dadda or Wallace tree multipliers (see [22] and [23]). Also, the bitwidth of the higher and lower parts are usually considered the same, i.e., $n/2$. Using this structure will allow us to apply VOS at a coarse level compared to state-of-the-art works that applied VOS at the fine level (see [3]), which may result in higher efficiency and lower overheads.

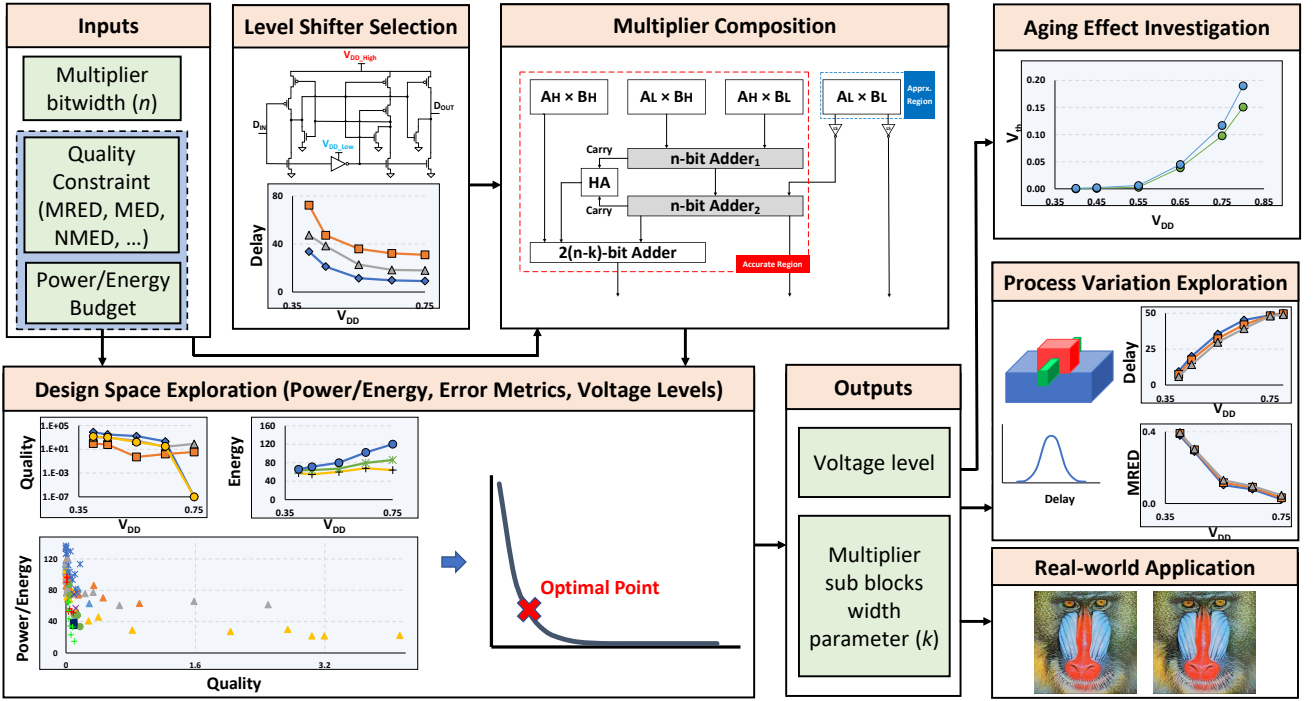

Fig. 3. system overview diagram of the proposed BL-VOS multiplier.

In [42], the coefficient partitioning technique was proposed to parallelize coefficient multiplications commonly used in DSP applications. In this method, the original coefficients are scaled to enable the partitioning of the multiplication into a smaller multiplication with shorter critical paths. Also, these smaller blocks can compute in parallel, and therefore, may allow for the supply voltage reduction.

In [43], a number of power-efficient multipliers, such as partitioned multipliers that break multiplier and multiplicand bits into two high significance and low significance parts, have been used to evaluate the performance of a low power programmable FIR filter. Specifically, using the partitioned multipliers has improved the area, power, and speed of the FIR filter. Based on the methods proposed in [42] and [43], using the partitioned multipliers may offer significant power efficiency improvements in DSP applications. In our proposed method, we apply the VOS-based approximate computing paradigm in the multipliers composed of smaller ones.

In FPGA-targeted works also, due to the availability of embedded multipliers with a given size on the FPGA devices, larger multipliers are composed by putting these smaller embedded multipliers together (see [44] and [45]).

### 3.2. Aging Factors and their Effects on the Performance

As the technology node shrinks in the nanoscale era, aging becomes more aggressive and undermines the system's reliability [19]. Generally, aging occurs due to some phenomena at the transistor level, such as Bias Temperature Instability (BTI), Hot Carrier Injection (HCI), Time-Dependent Dielectric Breakdown (TDDB), and Electromigration (EM) [19]. These factors cause undesired changes in the transistor characteristics and consequently reduce the performance and lifetime of the circuits. Specifically, HCI and BTI increase the threshold voltage ($V_{th}$) that lowers the performance. It should be mentioned that timing violation due to circuit performance reduction may cause some errors at the output. Therefore, compensating techniques are required to mitigate the aging effects. Fortunately, the effects of all the aforementioned aging mechanisms can be mitigated by reducing the supply voltage of the circuit [3]. In this paper, we study the BTI effects on the threshold voltage to measure the aging effect on the VOS-based multipliers.

Based on [20], the BTI phenomenon may affect both NMOS and PMOS transistors by creating traps at the Si/SiO$_2$ interface, where the effect of these traps on the threshold voltage change is obtained by [21]

$$\Delta V_{\text{th,BTI}} \cong A \, e^{-\frac{\kappa}{\theta}} t^\alpha \, E_{OX}^\gamma \, f^\beta \tag{1}$$

where $\theta$ is temperature in kelvin (K), $t$ is the total stress time in seconds (s), and $f$ is the duty factor of the stress s-ignal. $A$, $\kappa$, $\alpha$, $\beta$, and $\gamma$ are technology-dependent parameters discussed in [3]. Moreover, in this equation, $E_{OX}$ is the electric field induced on the gate oxide and is defined by [21]

$$E_{OX} = (V_{DD} - V_{th})/T_{INV} \qquad (2)$$

where $T_{INV}$ is the inversion layer thickness. Based on (1) and (2), the aging mechanism is a strong function of the supply voltage level ($V_{DD}$). Thus, employing VOS can significantly reduce the aging effects on the arithmetic units.

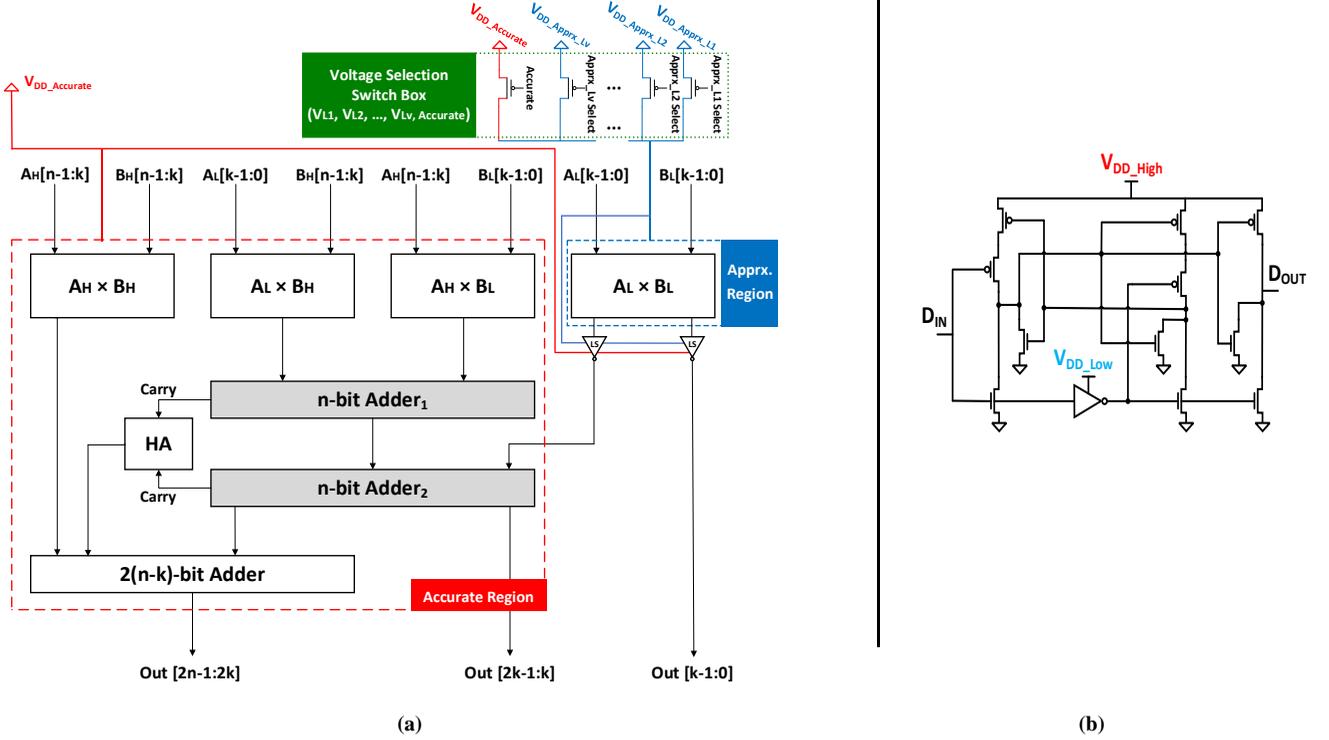

Fig. 4. (a) An example structure of the proposed accuracy configurable multiplier, in the case of $A_L \times B_L$ block is in the approximate voltage level and the other blocks are in the accurate voltage levels, (b) the circuit diagram of the level shifter (LS) [24].

## 4. THE PROPOSED GENERIC ACCURACY CONFIGURABLE MULTIPLIER

In this section, first, the structure of the proposed generic accuracy configurable multiplier is introduced. Next, the details of applying the VOS to the components of the proposed approximate multiplier are studied. The system overview diagram of the proposed BL-VOS multiplier, including the design flow and evaluation steps, is shown in Fig. 3. In this flow, based on the user-defined inputs including the multiplier bit width ($n$), quality constraints, and power/energy budget, the design space is explored to select the optimal structure of the proposed BL-VOS multiplier to meet those constraints/budget. The outputs of this flow are the size and voltage level of multiplier sub-blocks. Based on the selected voltage levels for the sub-blocks of the BL-VOS multiplier, it is necessary to use level shifters at the output of those sub-blocks that drive the components with higher voltage levels compared to their voltage levels. Afterward, some simulations are performed to investigate the effects of aging and process variation on the different structures of the proposed BL-VOS multiplier. Finally, the efficacy of the Bl-VOS multiplier is assessed for real-world applications such as image processing applications.

One of the main challenges in state-of-the-art VOS-based approximate multipliers is the complexity of controlling the voltage levels of the multiplier components. As an example, in the work proposed in [3], there are many compressors, half-adders, and full-adders that may operate at different voltage levels (see Fig. 1. b), which requires several controlling signals to adjust the voltage levels of each component. Such a fine granularity makes the decision more complex for the application/software that may employ several instantiates of that multiplier since it may require a very complex algorithm to select the voltage level of each component inside all the multipliers required for implementing that application. However, our proposed BL-VOS multiplier is designed in a coarse granularity (block level). Thus, fewer voltage-controlling signals are required compared to the state-of-the-art works, where there are only four sub-blocks inside a BL-VOS multiplier. This advantage is also useful for the application/software that employs the Bl-VOS multiplier since it requires a simpler algorithm for selecting the voltage level of the instantiated BL-VOS multipliers, i.e., there is a lower number of independent variables (voltage levels of components) to achieve the minimum power/energy consumption of the application.

Furthermore, a lower number of level shifters is required for composing the BL-VOS multiplier, rather than the fine-grained multiplier of [3]. As will be discussed in Section 5.4, besides the aforementioned ad-vantages, our proposed BL-VOS multiplier may achieve higher energy efficacy and lower delay for a given quality constraint, compared to the fine-grained VOS-based multipliers. In the following, we discuss the structure and design steps of the proposed BL-VOS multiplier.

### 4.1. The Proposed Structure

Fig. 4.a shows the overall structure of the proposed VOS-based accuracy configurable multiplier. Compared to the general structure of a block-based multiplier (see Fig. 2), to provide a larger design space resulting in different levels of accuracies and various energy consumption reductions, we consider an asymmetric width for the higher and lower parts of the multiplier, i.e., $A_H$, $B_H$, $A_L$, and $B_L$. This modification is achieved by defining the parameter $k$, the bitwidth of the input lower parts, including $A_L$ and $B_L$. Therefore, the bitwidth of higher and lower parts is equal to $[n-1:k]$ and $[k-1:0]$, respectively, where $0 < k < n$.

As shown in Fig. 4.a, two $n$-bit adders (the gray shaded blocks), one $2(n-k)$-bit adder, and one HA are employed to generate the $2n$-bit output of the multiplier. Now, depending on the different supply voltage levels applied to the four multiplier sub-blocks, including $A_L \times B_L$, $A_H \times B_L$, $A_L \times B_H$, and $A_H \times B_H$, various accuracy levels can be supported, with different amount of energy savings. In Fig. 4.a, the sub-blocks whose voltage level is kept at the nominal level (i.e., accurate voltage level) are enclosed in the red dashed boundary. The region inside the red boundary is referred as the accurate region. Also, the blocks with the lower voltage level (approximate voltage level) are separated by the blue dash line as the approximate region. Moreover, the voltage level of the approximate region is selected by the *Apprx_Li select* signals, where $1 \leq i \leq v$ ($v$ is the number of voltage levels). These signals are driven by the *Voltage Selection Switch Box* highlighted by the green dotted line in Fig. 4.a. To support the accurate operating mode by the BL-VOS-based multiplier, we also provided the accurate voltage level for the *Voltage Selection Switch Box*. Thus, by enabling this voltage level, the proposed multiplier is switched to the accurate operating mode, on the fly, which can be used in those states that accurate multiplication is required.

It is important to note that, to retain the performance, level shifters are required at those input signals of the components in the accurate region that are driven by the approximate region [24]. These level shifters convert the voltage of outputs of the approximate region to the accurate voltage level. However, depending on the design constraints, different level shifters can be used in our proposed multiplier, e.g., see the level shifters proposed in [24], [25], and [26]. Fig. 5 shows the delay of these level shifters for the different approximate voltage levels. The experimental setup to obtain these results will be discussed in Subsection 5.1. Also, to minimize the area and power consumption overheads, we used the minimum size transistors in these level shifters.

As shown in Fig. 5, the level shifter proposed in [24] offers the lowest delay compared to the other investigated level shifters. Specifically, for the voltage levels from 0.55V to 0.75V the imposed delay is negligible (~10 ps). Thus, we use this level shifter for the different structures of our proposed BL-VOS multiplier. The circuit diagram of this level shifter is shown in Fig. 4.b.

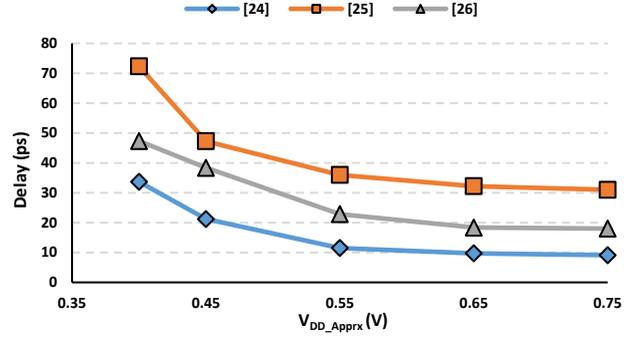

Fig. 5. Delay of the studied level shifters for the different approximate voltage levels.

### 4.2. Alpha Power Law and Applying the VOS Technique

Based on the alpha power law, the circuit delay is modeled by [3]

$$\text{Task}_{\text{Delay}} \propto \frac{V_{DD}}{(V_{DD} - V_{th})^\alpha} \quad (3)$$

where $\alpha$ is a technology-dependent parameter that is considered 1.3 for the technologies below 20-nm [27]. Based on (3), a lower voltage level ($V_{DD}$) induces a longer delay. Therefore, applying the VOS technique may increase the delay of the circuit and violate the timing constraints, where this violation may result in some errors at the circuit output. This is because of the changes in the timing of paths of the circuit [28]. Moreover, the lower supply voltage imposes more timing violations, and consequently, more errors occur at the output.

In our proposed structure, the amount of energy saving is proportional to the number of components that VOS is applied to them. However, applying VOS to the less important blocks (e.g., $A_L \times B_L$) may significantly improve the lifetime with a small error. Also, voltage overscaling of the more important blocks (e.g., $A_H \times B_H$) may drastically increase the error. Thus, we used the Bi-VOS method presented in [6] and [7] in our proposed structure, where the nominal voltage level that offers accurate results can be employed for the more important blocks.

As explained in Subsection 4.1, based on applying the VOS technique to the different components of the proposed structure (see Fig. 4.a), different levels of energy consumption reduction are provided, as well as various numbers of accuracy levels. TABLE II shows the different structures of the proposed multiplier, including the BL-VOS$_0$ to BL-VOS$_4$, where applying the VOS technique to the different components results in the different structures. Among these structures, the BL-VOS$_0$ corresponds to the accurate multiplier, where the voltage level of all components is set to the accurate one. Such a partitioning requires fewer level shifters compared to the fine-granularity structures where it is possible to group partial product bits by weight. This also lets to switch between approximate and accurate operating modes where all the voltage of all sub-blocks is switched to nominal one.

TABLE II THE DIFFERENT STRUCTURES OF THE PROPOSED MULTIPLIER BY APPLYING THE VOS TECHNIQUE TO THE DIFFERENT COMPONENTS

| Structure | $A_L \times B_L$ | $A_H \times B_L$ | $A_L \times B_H$ | $A_H \times B_H$ | $n$-bit Adder$_1$ | $n$-bit Adder$_2$ | $2(n-k)$-bit Adder | HA |
|---|---|---|---|---|---|---|---|---|
| BL-VOS$_0$ | - | - | - | - | - | - | - | - |
| BL-VOS$_1$ | ✓* | - | - | - | - | - | - | - |
| BL-VOS$_2$ | ✓ | ✓** | - | - | - | - | - | - |
| BL-VOS$_3$ | ✓ | ✓ | ✓ | - | ✓ | - | - | - |
| BL-VOS$_4$ | ✓ | ✓ | ✓ | ✓ | ✓ | ✓ | ✓ | ✓ |

\* The voltage of ticked blocks is overscaled.

\*\* Since the weight of the $A_H \times B_L$ and $A_L \times B_H$ blocks are the same, for the BL-VOS$_2$ structure, only the voltage of one of these blocks should be overscaled, arbitrarily (up to the designer).

Note that the parameter $k$ is yet another knob that results in a larger design space. Therefore, the more appropriate structure of the proposed multiplier is composable for the different levels of QoS requirement for various applications, in terms of quality and energy efficiency.

### 4.3. Hardware Complexity of the Proposed Accuracy Configurable Multiplier

Besides the achievable energy efficiency offered by the proposed generic accuracy configurable multiplier, there are some hardware complexities induced by such a design, e.g., number of level shifters, number of voltage levels, and method to determine which of those five possible structures should be employed in a given application. These challenges are discussed in the following.

The first issue is the number of level shifters that increase the hardware complexity. As discussed in Subsection 4.1 and shown in Fig. 4.a, level shifters are required to drive a block that operates at a high voltage with the blocks with lower voltages. Specifically, depending on the BL-VOS bit width parameters ($n$, $k$), for the BL-VOS$_1$, BL-VOS$_2$, and BL-VOS$_3$ structures, $2k$, $2k + n$, and $2k + n + 1$ number of level shifters are required, respectively. Also, there is no need for level shifters in the BL-VOS$_4$ structure since its all components are operates at low voltage levels.

Another challenge is providing continuous supply voltage for VOS-based units, which may impose significant implementation overheads. However, using a few discrete voltages is good enough to achieve significant energy savings, and is relatively simpler from an implementation perspective [35]. Therefore, similar to [3], in this paper, we used five different discrete approximate voltages levels. Among these five proposed BL-VOS structures, the designer is responsible to determine which one should be employed for the implementation. However, some recent works such as [36] and [37] have proposed methods to determine the required accuracy level of the approximate arithmetic units for the different applications. Thus, depending on the required accuracy level for a given approximate multiplier, one of our five proposed structures could be employed in the implementations.

Furthermore, at the architecture level, where several accuracy configurable arithmetic units may be employed in a quality configurable architecture, an additional unit is required to activate the accuracy control signals of the different units. As an example, in [38], an accuracy management process (AMP) was proposed to control the accuracy of the approximate arithmetic units employed in a quality adjustable CGRA. Therefore, the accuracy management unit at the higher architecture level will activate the enable signals of the *Voltage Selection Switch Box* of the proposed approximate multiplier during the runtime based on a given accuracy management policy.

Note that, another method for building a multiplier with the smaller ones is beyond four blocks, where the number of blocks increased with the multiplier size increment [39]. However, in this case, the number of level shifters, control signals, and the size of the *Voltage Selection Switch Box* can increase significantly which may lower the gains achieved by the VOS. Also, in this case, due to the considerable enlarging of the design space, finding the optimum structure would become more difficult.

Finally, to reduce the total area and hardware complexity of the circuit, some complementary methods such as least significant bits truncation may be used, e.g., in the X-Dadda multiplier [3], a 4-bit truncation method was used in the first four LSBs of the Dadda multiplier. Without loss of generality, such a complementary method can be used in our proposed BL-VOS structures for the area efficiency, e.g., one may truncate some bits of the $A_L \times B_L$ block. Note that, although the truncation method reduces the area and hardware complexity, it may greatly increase the error. In the Subsection 5.5, we investigate the effect of the truncation method on our proposed BL-VOS multiplier.

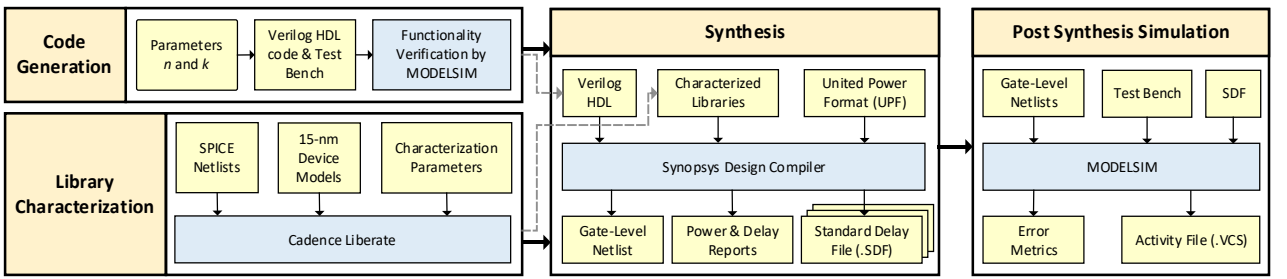

Fig. 6. Design steps and simulation tool flow used for the synthesis and error analysis of the proposed BL-VOS-based approximate multiplier.

## 5. RESULTS AND DISCUSSION

This section presents simulation flow, design steps, and obtained results for the different structures of the proposed BL-VOS-based multiplier.

### 5.1. Simulation Flow

Fig. 6 shows details of the design steps and simulation tool flow used for error analysis and evaluating the proposed approximate multiplier, including code generation, library characterization, synthesis, and post-synthesis simulation steps. At first, the Verilog HDL codes for the different structures of the BL-VOS presented in TABLE II, are generated. To evaluate the efficiency of the proposed concept, a 15-nm FinFET technology [29] with a nominal operating voltage of 0.8V was used for synthesizing all the investigated multipliers, including the different structures of our proposed BL-VOS multiplier and the designs of [3], [11], [12], [13], [17], and [18].

In this paper, we used five different approximate voltages levels, including 0.4, 0.45, 0.55, 0.65, and 0.75V for approximate blocks, where each corresponding library to each voltage level was characterized using the Cadence Liberate tool. Level shifter cells also were added to the characterized libraries. It should be noted that since we used FinFET devices, the number of the fins determines the area and delay overheads of the switches. In this article, the number of fins for the voltages of 0.4, 0.45, 0.55, 0.65, and 0.75V were chosen 37, 27, 19, 14, and 12, respectively [3].

Next, the verified HDL codes, the united power format (UPF) file, and the characterized libraries for the given approximate voltage levels are fed to the Synopsys Design Compiler (DC). Note that, the UPF file that is the standard format of the low- power design [30], is necessary due to using the different voltages levels in our proposed multipliers. Next, at the post-synthesis simulation step, the gate-level netlists generated by the DC along with the Standard Delay File (SDF) are used to calculate the error induced by applying the VOS. It should be noted that the SDF file for each design is generated by the DC, in which the delay of the various logic gates is provided [3]. TABLE III shows the delay of the BL-VOS multiplier when operating at the accurate voltage (i.e., $V_{DD}$ = 0.8V), for the different values of $k$. As discussed before, timing violations may cause errors in the outputs of the circuit. In our simulations, the obtained delays of the accurate BL-VOS structures (operating at the accurate voltage, i.e., $V_{DD}$ = 0.8V) under the different values of $k$ (see TABLE III), were used for setting the clock frequency of the flip-flops that sampled the BL-VOS outputs. By lowering the approximate voltage ($V_{DD\_Apprx}$), these timing requirements are violated and may cause errors in the circuit output. Thus, similar to [3], there is no need to consider specific timing requirements for the flip-flops used with the proposed multiplier. In the next subsection, the common error metrics in the approximate computing literature are introduced.

### 5.2. Error Analysis

To evaluate the accuracy of the approximate designs, sev-eral error metrics were proposed in the approximate computing literature, such as error rate (ER), error distance (ED), mean error distance (MED), mean relative error distance (MRED), and normalized mean error distance (NMED) [9]. TABLE IV shows the details of these error metrics. In the application/software level, mean and variance of error can be employed to calculate the error of approximate arithmetic units that may generate indeterministic errors, at the output of application/software that used those approximate units. Similarly, since the BL-VOS result in an indeterministic error, one may calculate the mean and variance of error of the different structures of the proposed BL-VOS to use them in the application/software level.

TABLE III  DELAY OF THE BL-VOS MULTIPLIER UNDER THE DIFFERENT VALUES OF $k$

| | Delay (ps) | | | | | |
|---|---|---|---|---|---|---|
| | 8-bit | | | 16-bit | | |
| Structure | $k = 2$ | $k = 4$ | $k = 6$ | $k = 4$ | $k = 8$ | $k = 12$ |
| BL-VOS | 91.57 | 102.35 | 112.82 | 152.33 | 172.62 | 211.06 |

For the error analysis, in this work, we examined 8, 16-bit BL-VOS structures. Also, we considered $k$ = 2, 4, and 6 for the 8-bit multiplier, and $k$ = 4, 8, and 12 for the 16-bit multiplier. Fig. 7 (a)-(i) show the MED, MRED, and NMED of the different structures of the 8-bit BL-VOS multipliers under the different voltage levels. The results were obtained by applying 10,000 (1,000,000) uniform random inputs in the case of 8-bit (16-bit) BL-VOS multipliers. As expected, the values of MED, MRED, and NMED are increasing with the increment of the number of approximate components (from BL-VOS$_1$ to BL-VOS$_4$) and the parameter $k$.

Also, among the different studied 8-bit approximate multipliers, the maximum MED, MRED, and NMED belonged to the BL-VOS$_4$ structure for $k$ = 2 and $V_{DD\_Apprx}$ = 0.4V. Because, as shown in TABLE II, in the BL-VOS$_4$, the voltage of all the components is overscaled. However, using more close voltages to the nominal voltage (0.8V) results in higher accuracy. For example, as shown in Fig. 7, at 0.75V, most BL-VOS structures have negligible error metrics. But, in a few points, the mentioned trend is not followed the same as expected. For example, see the NMED of the BL-VOS$_1$ in Fig. 7.h, where the NMED at 0.55V (see pointer ①) is lower than the NMED at 0.65V (see pointer ②). Such cases occur due to the various gates with the different sizes (delays) employed by the Synopsys Design Compiler to synthesize the BL-VOS structures at the different voltage levels independently [3]. Fig. 7 (j)-(r) show the MED, MRED, and NMED results of the 16-bit approximate BL-VOS multiplier for the various $k$ and $V_{DD\_Apprx}$. Similar to the results of the 8-bit multipliers, the BL-VOS$_1$ (BL-VOS$_4$) structure shows the best (worst) results.

Finally, in (4), which is shown at the top of the next page, we proposed a formula to calculate the sensitivity of error metrics ($S_{Err\_Metric}$), i.e., MED, MRED, NMED, to the changes of approximate knobs (voltage level and parameter $k$) for the different structures of the BL-VOS multiplier(see TABLE II). The parameters involved in (4), are

$$S_{Err\_Metric}(BL-VOS_i, k, V_{DD\_Apprx\_Lv}) = \frac{Err\_Metric(BL-VOS_i, k, V_{DD\_Apprx\_Lv+1}) - Err\_Metric(BL-VOS_i, k, V_{DD\_Apprx\_Lv})}{V_{DD\_Apprx\_Lv} - V_{DD\_Apprx\_Lv+1}} \quad (4)$$

TABLE IV COMMON ERROR METRICS IN THE APPROXIMATE COMPUTING LITERATURE

| Error Metric | Definition | Description |
|---|---|---|
| ER | $\frac{\#errorneous\ outputs}{total\ number\ of\ outputs}$ | - |
| ED | $|O - O'|$ | $O$: Exact output $O'$: Apprx. output |
| MED | $\frac{1}{2^{2n}}\sum_{i=1}^{2^{2n}}|ED_i|$ | $n$: Bit width of input operands |
| MRED | $\frac{1}{2^{2n}}\sum_{i=1}^{2^{2n}}\frac{|ED_i|}{O_i}$ | - |
| NMED | $\frac{MED}{D} = \frac{1}{2^{2n}}\sum_{i=1}^{2^{2n}}\frac{|ED_i|}{D}$ | $D$: Maximum possible error |

the type of multiplier structure (BL-VOS$_i$), parameter $k$, and the approximate voltage level ($V_{DD\_Apprx\_Lv}$). Note that, the index $i$ ($1 \leq i \leq 4$) is used to represent different multiplier structures in the set of {BL-VOS$_1$, BL-VOS$_2$, BL-VOS$_3$, BL-VOS$_4$}, the index $k$ ($0 < k < n$) is the width parameter of multiplier sub-blocks (see Subsection 4.1), and the index $v$ ($0 \leq v < 4$) is used to indicate the different voltage levels in the set of {0.75, 0.65, 0.55, 0.45, and 0.4V}.

According to the obtained results for the MRED study, for all values of $k$, when the voltage level decreases from 0.75V to 0.45V, $S_{Err\_MRED}$ increases, on average, by 76×, 118×, 168×, and 180× for the 8-bit BL-VOS$_1$, BL-VOS$_2$, BL-VOS$_3$, and BL-VOS$_4$ structures, respectively. Therefore, the structures with more voltage overscaled blocks result in higher $S_{Err\_MRED}$. Also, the changes of the $S_{Err\_MRED}$ with increasing $k$ were examined. Based on the obtained results, for all considered values of $V_{DD\_Apprx\_Lv}$, when the $k$ is increased from 2 to 6, $S_{Err\_MRED}$ increases on average by 2×, 11×, 38×, and 3× for the 8-bit BL-VOS$_1$, BL-VOS$_2$, BL-VOS$_3$, and BL-VOS$_4$ structures, respectively. Thus, the $S_{Err\_MRED}$ is enlarged as the parameter $k$ increases.

### 5.3. Design Parameters Evaluation

Fig. 8 (a)-(d) show the energy consumption of the different structures of the proposed 8-bit BL-VOS multiplier. As the results show, the structures with $k = 4$ and $k = 6$ offer more energy consumption, on average, 6.5% and 9.8%, compared to the $k = 2$, respectively. Specifically, the maximum amount of energy reduction (81.6%) is related to the BL-VOS$_4$ structure with $k = 2$ and $V_{DD\_Apprx} = 0.4V$ (see pointer ③ in Fig. 8.d).

For the BL-VOS$_1$, although a few numbers of level shifters are required, fewer approximate components result in lower energy efficacy compared to the other structures. Also, as shown in Fig. 5, level shifters at closer voltages to the nominal one result in lower delay, i.e., the transfer of a very low voltage (e.g., 0.4V) to the 0.8V consumes more energy than the transfer of a higher voltage (e.g., 0.75V) to the 0.8V. Thus, the BL-VOS$_1$ at 0.4V and $k = 2$ (see pointer ④ in Fig. 8.a) consumes more energy than the BL-VOS$_1$ at 0.75V and $k = 2$ (see pointer ⑤ in Fig. 8.a). Generally, for the different structures of the 8-bit BL-VOS multipliers, BL-VOS$_4$ (BL-VOS$_1$) shows the best (worst) energy consumption compared to the other structures.

Finally, the energy consumption results of 16-bit BL-VOS multipliers are shown in Fig. 8 (e)-(h). Based on the results, the structures with $k = 4$ offer the best energy efficacy compared to other structures. Also, the maximum energy consumption reduction (~82.4%) belongs to the BL-VOS$_4$ structure for $k = 8$ and $V_{DD\_Apprx} = 0.4V$ (see pointer ⑥ in Fig. 8.h).

### 5.4. Design Parameters Comparisons

In this subsection, the power and energy vs. MRED comparison results of our proposed BL-VOS-based multiplier with state-of-the-art approximate multipliers (see TABLE I) are presented. As mentioned before, the delay of VOS-based approximate multipliers is considered equal to the delay of the corresponding exact structure. Thus, the comparison of the delay parameter between our proposed structures and other approximate multipliers have not been considered.

The results of power vs. MRED for different studied approximate multipliers are shown in Fig. 9. Based on the results, our proposed BL-VOS-based structures create the most appropriate trade-offs between power consumption and MRED. In particular, the best result belongs to the BL_VOS$_2$ with $k = 2$ and $V_{DD\_Apprx} = 0.75V$ (see pointer ⑦ in Fig. 9), where the power consumption is 97.815 mW, and MRED is ~1E-09. Also, the lowest power consumption is related to the YUS-V2, where power is 14.713 mW, and MRED is 0.1 (see pointer ⑧ in Fig. 9).

Energy vs. MRED results for different studied approximate multipliers is plotted in Fig. 10. Based on the results, our proposed BL-VOS-based multipliers offer the most suitable trade-offs. Especially, the best result is related to the BL-VOS$_2$ structure with $k = 2$ and $V_{DD\_Apprx} = 0.75V$ (see pointer ⑨ in Fig. 10), where the energy consumption is 8.96 pJ, and MRED is ~1E-09. These achieved energy efficiencies are mainly due to employing the VOS at the different blocks of the proposed BL-VOS multiplier, rather than an exact multiplier that operates in the nominal voltage of the given technology. Also, the lowest energy consumption among the all studied approximate multipliers is related to the YUS-V2 multiplier with energy consumption of 0.32 pJ, and MRED about 0.1 (see pointer ⑩ in Fig. 10). In TABLE V, we have shown the points belonging to the proposed BL-VOS and the X-Dadda of [3], in the four ranges of MRED, from 0 to 0.16. Note that in this table, for a given structure and MRED range, when there are different points their MRED values are the same, we only brought the point with the minimum energy consumption.

Based on the results demonstrated in Fig. 9 and Fig. 10, with lowering the voltage level, the power and energy consumption are decreased, whereas the error parameters are increased. TABLE VI shows the design indexes of the BL-VOS and the state-of-the-art VOS-based X-Dadda [3] under the different studied values of $V_{DD}$.

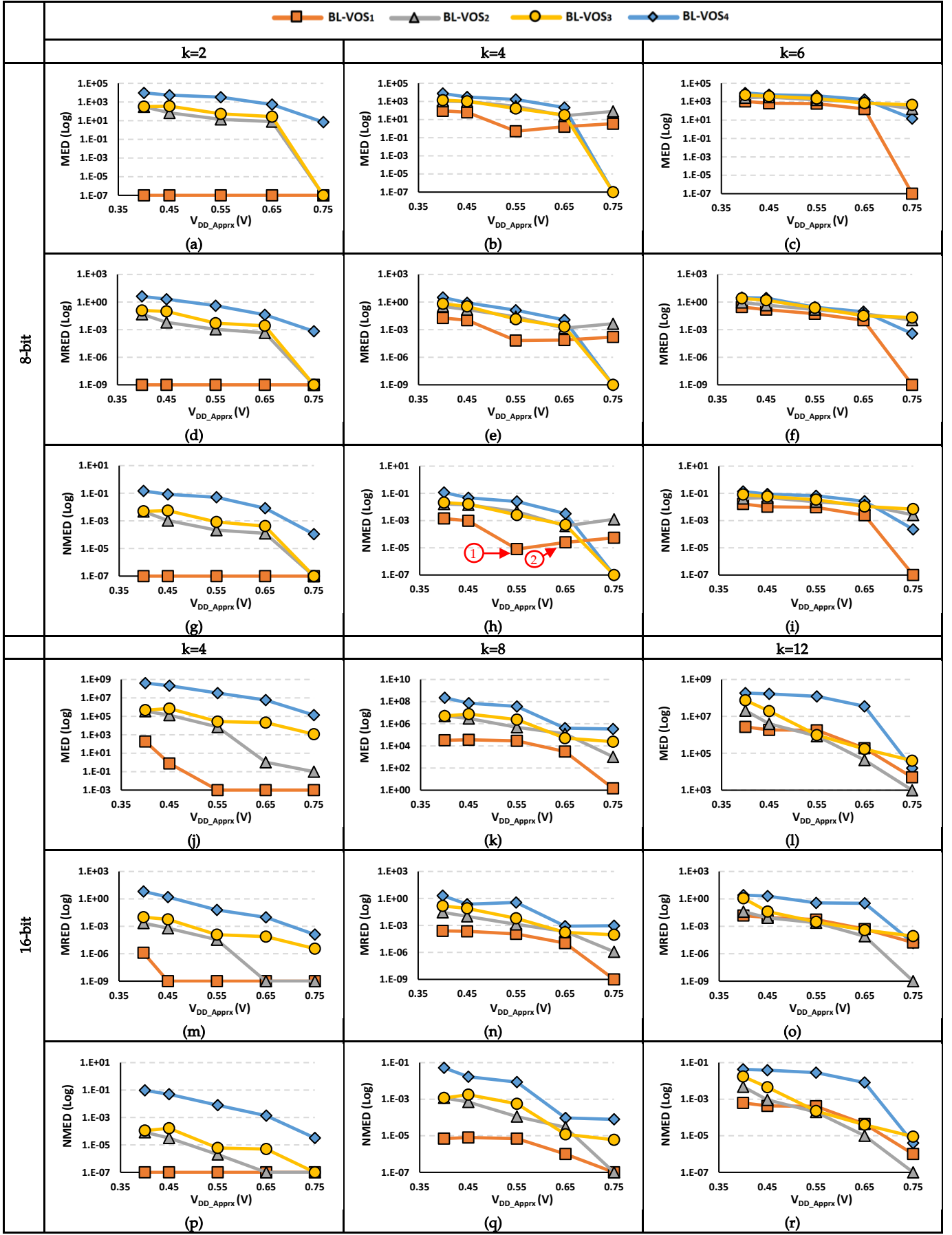

Fig. 7. MED, MRED, and NMED of different structures of the BL-VOS multiplier under the different values of k and $V_{DD\_Apprx}$, (a)-(i) the 8-bit multipliers, (j)-(r) the 16-bit multipliers.

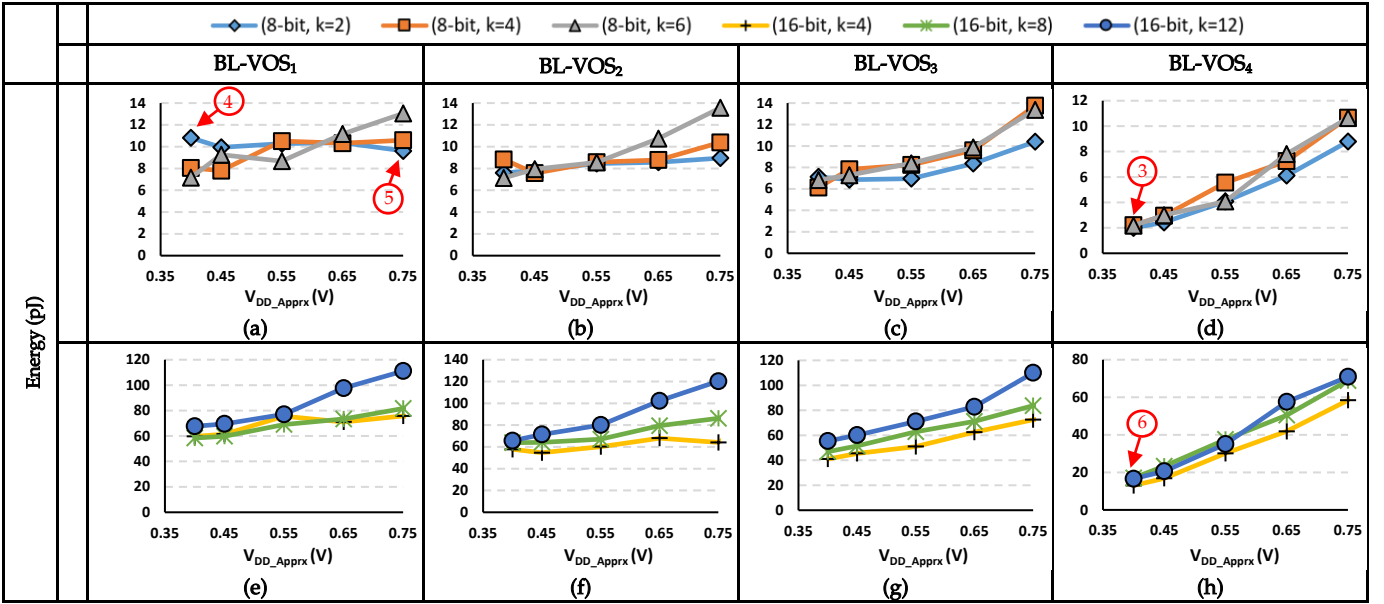

Fig. 8. Energy consumption of different structures of the BL-VOS multiplier under the different values of k and $V_{DD\_Apprx}$, (a)-(d) the 8-bit multipliers, (e)-(h) the 16-bit multipliers.

### 5.5. Employing the Truncation Method in the BL-VOS Structures

In this subsection, we investigate the effect of employing the truncation method in the proposed BL-VOS multiplier. TABLE VII shows delay, energy, MRED, and NMED for the four different proposed structures of the BL-VOS, under the without truncation ($w/o\_t$) and with truncation ($w\_t$) methods. Please note that these results were obtained for $n = 16$, $k = 8$, and $V_{DD\_Apprx} = 0.45V$. The results metrics were improved, on average, by 10.5% and 25.8% compared to the $w/o\_t$ structures, respectively. Also, the highest delay and energy reductions were 14.54% and indicate that by employing the truncation method in the four different structures of the BL-VOS, delay and energy 37.7% correspond to BL-VOS$_1$ and BL-VOS$_3$ structures, respectively. However, MRED and NMED of the $w\_t$ BL-VOS structures were worsened, on average, by 49.1%, and 11.1%, respectively.

Note that, the accuracy metrics of the $w\_t$ structures may be affected for two reasons. First, decreasing the delay in the $w\_t$ structures may increase the accuracy since a shorter time is required to generate the output. Second, employing the truncation method will increase error due to omitting some LSBs in a given structure. Thus, in some cases, e.g., the BL-VOS$_2$, by using the truncation method, the MRED and NMED metrics were improved, by 7.8% and 14.3%, respectively. Besides the truncation method, one may gate the input data of the different sub-blocks of the BL-VOS multiplier. In this case, the voltage of the unused sub-blocks could be power gated to reduce the energy consumption, while resulting in a deterministic error rather than the voltage overscaling. The simulation results of this case have been presented in TABLE VII entitled the power gating ($PG$) mode. It is trivial that only the first three blocks of the multiplier could be turned off, and thus, the $A_H \times B_H$ block is not considered for the power gating in TABLE VII.

TABLE V [ENERGY CONSUMPTION, MRED] OF THE BL-VOS AND X-DADDA MULTIPLIERS, IN THE FOUR DIFFERENT RANGES OF MRED

| Range of MRED | [Energy (pJ), MRED] | | | |
|---|---|---|---|---|
| | (0, 0.04) | (0.04, 0.08) | (0.08, 0.12) | (0.12, 0.16) |
| BL-VOS$_1$ | [9.62, 0]<br>[7.80, 0.01]<br>[8.05, 0.02] | [8.67, 0.05] | - | [9.27, 0.15] |
| BL-VOS$_2$ | [7.81, 0]<br>[13.57, 0.01]<br>[8.58, 0.02] | [7.60, 0.05] | - | [7.57, 0.15] |
| BL-VOS$_3$ | [6.96, 0]<br>[8.24, 0.01]<br>[13.38, 0.02]<br>[9.87, 0.03] | - | [6.86, 0.10]<br>[7.13, 0.12] | - |
| BL-VOS$_4$ | [8.79, 0]<br>[7.25, 0.01] | [6.11, 0.04] | [7.81, 0.09] | [5.56, 0.13] |
| X-Dadda [3] | [9.58, 0]<br>[7.83, 0.02]<br>[8.35, 0.03] | [7.70, 0.04]<br>[9.69, 0.05]<br>[7.51, 0.06]<br>[6.98, 0.08] | [7.80, 0.11]<br>[6.34, 0.12] | [5.96, 0.14] |

### 5.6. Effects of Process Variation

PV can change the circuit characteristics in the nanoscale era, where lowering the voltage level may intensify this phenomenon. The reason is that the operating voltage of the devices is getting closer to the threshold level [31]. Here we study the effects of PV on the proposed BL-VOS multipliers. Specifically, we examine the PV effects on the different structures of the BL-VOS multiplier with $k = 4$, under two voltage levels of 0.4V and 0.55V. The considered parameters for the PV assessments are listed in TABLE VIII. Note that the local PV assessment has been

TABLE VI   Key design indexes of the 8-bit BL-VOS and X-Dadda approximate multipliers, under the different studied values of $V_{DD}$ @ $k = 4$

| Using a 15-nm FinFET technology | Delay (ps) | $V_{DD} = 0.4V$ | | | $V_{DD} = 0.45$ | | | $V_{DD} = 0.55$ | | | $V_{DD} = 0.65$ | | | $V_{DD} = 0.75$ | | |
|---|---|---|---|---|---|---|---|---|---|---|---|---|---|---|---|---|
| | | Energy (pJ) | MRED | NMED | Energy (pJ) | MRED | NMED | Energy (pJ) | MRED | NMED | Energy (pJ) | MRED | NMED | Energy (pJ) | MRED | NMED |
| BL-VOS$_1$ | 101 | 8.0 | 0.018 | 0.001 | 7.8 | 0.011 | 0.001 | 10.5 | 0.000 | 0.000 | 10.3 | 0.000 | 0.000 | 10.6 | 0.000 | 0.000 |
| BL-VOS$_2$ | | 8.8 | 0.342 | 0.016 | 7.6 | 0.150 | 0.014 | 8.6 | 0.025 | 0.004 | 8.8 | 0.001 | 0.000 | 10.4 | 0.004 | 0.001 |
| BL-VOS$_3$ | | 6.1 | 0.660 | 0.021 | 7.8 | 0.335 | 0.016 | 8.2 | 0.013 | 0.002 | 9.6 | 0.002 | 0.000 | 13.7 | 0.000 | 0.000 |
| BL-VOS$_4$ | | 2.2 | 3.194 | 0.116 | 2.9 | 0.819 | 0.047 | 5.6 | 0.129 | 0.025 | 7.2 | 0.011 | 0.003 | 10.7 | 0.000 | 0.000 |
| X_Dadda [3] | 75 | 8.5 | 0.173 | 0.017 | 7.0 | 0.081 | 0.011 | 7.7 | 0.042 | 0.006 | 8.3 | 0.028 | 0.002 | 9.7 | 0.055 | 0.007 |

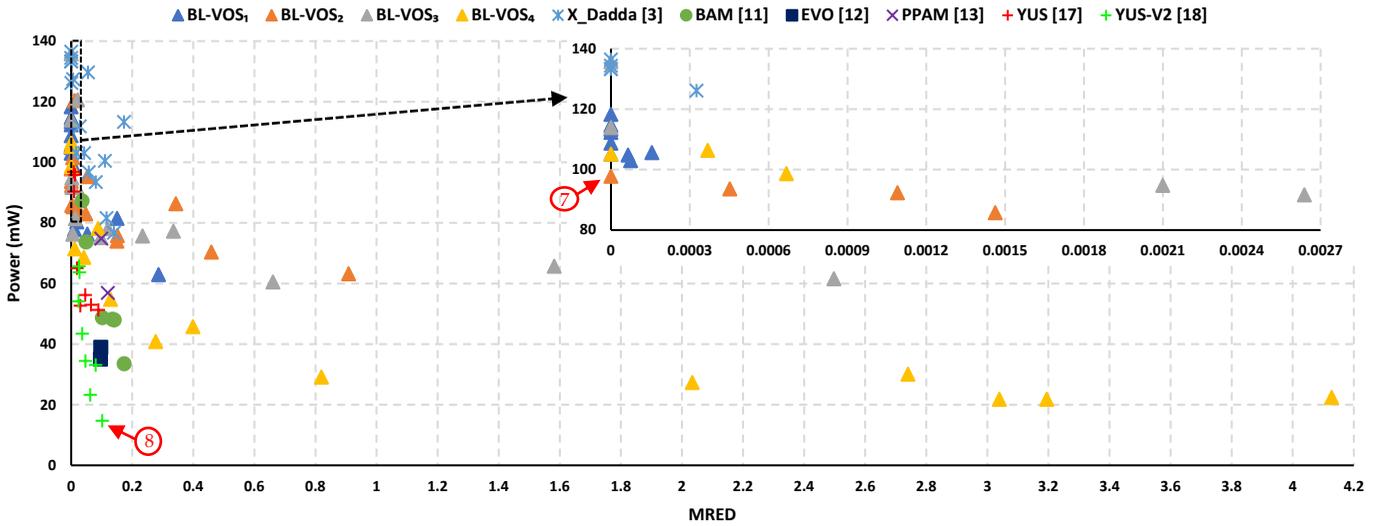

Fig. 9. Power vs. MRED for the different studied approximate multipliers.

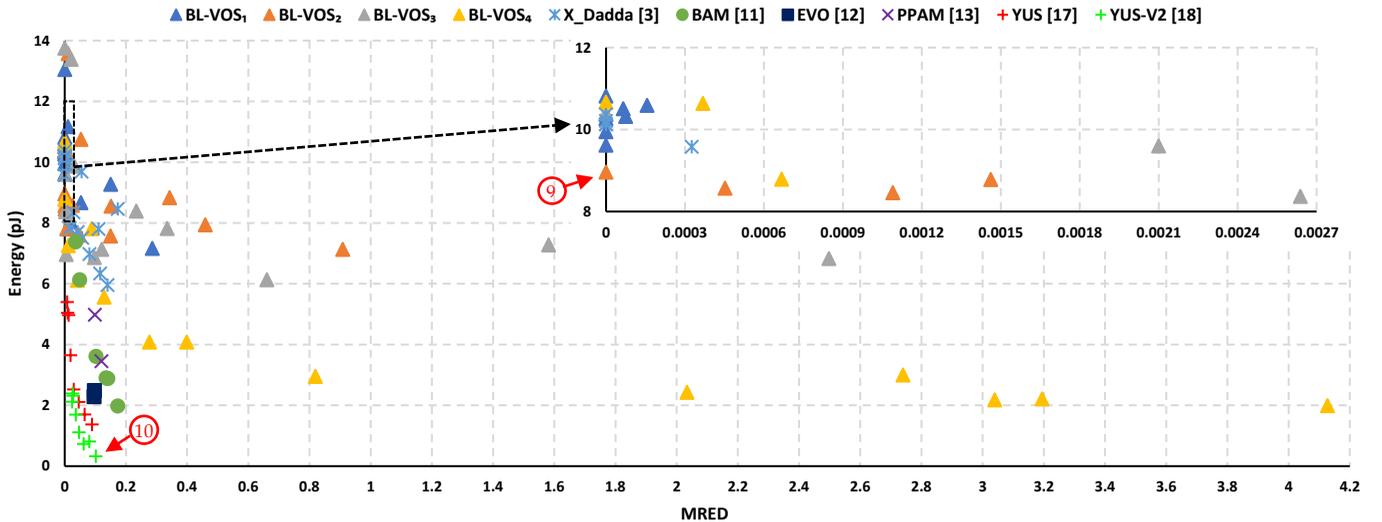

Fig. 10. Energy vs. MRED for the different investigated approximate multipliers.

TABLE VII   Design metrics of the different structures of the BL-VOS multiplier in the cases of without truncation ($w/o\_t$), with truncation ($w\_t$), and power gating (PG) modes, for $n = 16$, $k = 8$, $V_{DD\_APPRX} = 0.45V$, and $V_{DD} = 0.8V$

| Structure | Delay (ps) | | | Energy (pJ) | | | MRED | | | NMED | | |
|---|---|---|---|---|---|---|---|---|---|---|---|---|
| | $w/o\_t$ | $w\_t$ | PG | $w/o\_t$ | $w\_t$ | PG | $w/o\_t$ | $w\_t$ | PG | $w/o\_t$ | $w\_t$ | PG |
| BL-VOS$_1$ | 163.0 | 139.3 | 143.3 | 59.8 | 58.0 | 52.2 | 0.0002 | 0.0002 | 0.0002 | 0.0000 | 0.0000 | 0.0000 |
| BL-VOS$_2$ | 172.6 | 163.7 | 116.9 | 64.1 | 47.9 | 32.3 | 0.0103 | 0.0095 | 0.0647 | 0.0007 | 0.0006 | 0.0007 |
| BL-VOS$_3$ | 162.5 | 140.9 | 73.3 | 51.5 | 32.1 | 10.6 | 0.0792 | 0.0471 | 0.1206 | 0.0018 | 0.0010 | 0.0010 |
| BL-VOS$_4$ | 151.0 | 137.4 | - | 23.3 | 14.6 | - | 0.2266 | 0.7814 | - | 0.0169 | 0.0343 | - |

TABLE VIII THE PARAMETERS OF PV ASSESSMENTS

| Parameter | Description | Value |
|---|---|---|
| $L_g$ | The gate length | $3\sigma = 10\%$* |
| $t_{si}$ | The fin width | $3\sigma = 10\%$ |
| $H_{fin}$ | The fin height | $3\sigma = 10\%$ |
| $t_{ox}$ | The gate oxide thickness | $3\sigma = 5\%$ |

* of the nominal value.

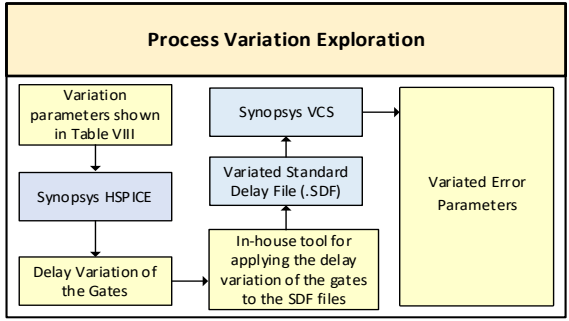

Fig. 11. Simulation flow to explore the PV effects on the different structures of the proposed BL-VOS approximate multiplier.

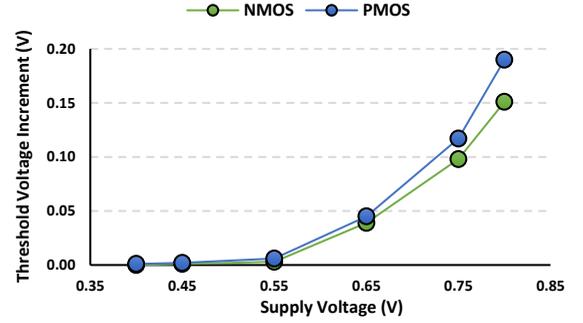

Fig. 12. Increment of NMOS and PMOS threshold voltages vs. supply voltage after ten years of operation.

considered for $t_{si}$ and $L_g$ due to line edge roughness (LER) [32].

Fig. 11 shows the simulation flow for exploring the PV effects on the different structures of the BL-VOS multiplier. Based on the parameters shown in TABLE VIII, we performed Monte Carlo simulations using the Synopsys HSPICE tool to extract the delay variations of the gates. Next, for applying the variations of the gates to the original SDF files, which are extracted by the DC tool, a developed in-house tool was used. In particular, 5,000 variated SDF files were generated for each design. Afterward, using the generated SDF files, each design was simulated by Synopsys VCS tool 5,000 times to calculate the changes of MED, MRED, and NMED parameters.

The variations of the error metrics under the PV impacts are shown in TABLE IX. Based on the results, for the BL-VOS$_1$, BL-VOS$_2$, and BL-VOS$_3$ structures, PV has a negligible effect on the investigated error metrics, on average, 0.78%, 0.40%, and 0.26%, respectively. Also, for the BL-VOS$_4$ structure, the maximum changes in the MED, MRED, and NMED are 0.8%, 4.4%, and 5.2%, respectively, which are belong to the 0.4V voltage level.

### 5.7. Aging Effects on the Proposed BL-VOS-based Approximate Multipliers

As discussed in Subsection 5.6, in the nanoscale era, lowering the operating voltage level may intensify the effects of the PV phenomenon on the circuit characteristics since the operating voltage of the devices is getting closer to the threshold voltage. Furthermore, in the approximate multipliers, we accept some minor errors to reduce the circuit's power and energy consumption. However, due to the aging, the resulting error may be out of the acceptable range. Therefore, studying the PV and aging effects on the accuracy of the approximate multipliers becomes more important. We used (1) to calculate changes of threshold voltage ($V_{th}$) of NMOS and PMOS transistors after ten years. At 0.4V and 0.8V, the initial threshold voltage of NMOS (PMOS) was considered 0.205

($-0.181V$) and 0.175 ($-0.190V$), respectively.

Fig. 12 shows the change of NMOS and PMOS threshold voltages vs. supply voltage after ten years of operation. Based on the results, for the 0.8V voltage level, the threshold voltages of NMOS (PMOS) increases by 0.151V (0.190V), while these parameters are about zero for the 0.4V voltage level.

The increase of the threshold voltage can greatly impair circuit performance and negatively affect circuit efficiency, and as shown in Fig. 12, this will get worse as the supply voltage increases. In other words, the aging effect is reduced by decreasing the voltage. Thus, employing the VOS can significantly reduce the aging effects on the VOS-based approximate arithmetic units. Especially, the delay increment due to the BTI can be calculated by (3). Fig. 13 shows the delay increment of the different structures of the 8-bit BL-VOS multiplier under the various values of $k$ and $V_{DD\_Apprx}$, after ten years of operation. As the results show, with increasing the number of approximate blocks and decreasing the approximate voltage level ($V_{DD\_Apprx}$), the delay changes are lowering compared to the exact voltage. Remarkably, the highest delay increment (~50%) being related to the BL-VOS$_1$ with $k = 2$ and at 0.8V (see pointer ⑪ in Fig. 13), and the lowest delay change (~1%) being related to the BL-VOS$_4$ with $k = 6$ and $V_{DD\_Apprx} = 0.4V$ (see pointer ⑫ in Fig. 13). However, delay increases lead to a higher error rate. To illustrate this, MRED reduction vs. $V_{DD\_Apprx}$ of approximate structures after ten years of operation are shown in Fig. 14. According to the results, the highest (lowest) MRED reduction is related to BL-VOS$_4$ (BL-VOS$_1$) with $k = 6$ ($k = 2$) and $V_{DD\_Apprx} = 0.4V$ (0.75V) (see pointer ⑬ and ⑭ in Fig. 14), which is equal to ~0.8 (~0.02).

### 5.8. Image Processing Applications

In this subsection, we examine the effectiveness of the different structures of our proposed BL-VOS-based approximate multipliers in two real-world applications, including image sharpening and smoothing applications [33]. For these image processing applications, we used five standard image benchmarks from [34]. Also, we used 8-bit BL-VOS multipliers with $k = 4$, under two voltages levels of 0.4V and 0.75V. TABLE X shows the mean structural similarity index metric (MSSIM) and energy reduction results of this study. Based on the results, the best MSSIM results belong to the structures with $V_{DD\_Apprx} = 0.75V$. Also, the maximum energy saving is 79.9% achiev-ed by the BL-VOS$_4$ in $V_{DD\_Apprx} = 0.4V$. Fig. 15 shows the output images of the investigated image processing applications for the "Baboon" image benchmark obtained by using 8-bit BL-VOS multipliers.

TABLE IX  VARIATION OF THE ERROR METRICS UNDER THE PV ASSESSMENTS

| Structure | | $k=4$, $V_{DD\_Apprx}=0.4V$ | | | $k=4$, $V_{DD\_Apprx}=0.55V$ | | |
|---|---|---|---|---|---|---|---|
| | | MED | MRED | NMED | MED | MRED | NMED |
| BL-VOS$_1$ | Nominal | 9.25E+01 | 1.84E-02 | 1.42E-03 | 5.12E-01 | 6.54E-05 | 8.00E-06 |
| | Mean | 9.25E+01 | 1.85E-02 | 1.42E-03 | 5.10E-01 | 6.36E-05 | 7.84E-06 |
| | Std | 1.58E+00 | 3.06E-04 | 2.40E-05 | 1.80E-02 | 2.12E-06 | 2.80E-07 |
| | Mean/Std | 59 | 60 | 59 | 28 | 30 | 28 |
| BL-VOS$_2$ | Nominal | 1.07E+03 | 3.42E-01 | 1.65E-02 | 2.97E+02 | 2.54E-02 | 4.57E-03 |
| | Mean | 1.08E+03 | 3.51E-01 | 1.68E-02 | 2.94E+02 | 2.51E-02 | 4.54E-03 |
| | Std | 1.10E+01 | 4.80E-03 | 1.71E-04 | 6.87E+00 | 5.08E-04 | 1.05E-04 |
| | Mean/Std | 98 | 73 | 98 | 43 | 49 | 43 |
| BL-VOS$_3$ | Nominal | 1.38E+03 | 6.60E-01 | 2.13E-02 | 1.66E+02 | 1.36E-02 | 2.56E-03 |
| | Mean | 1.38E+03 | 6.61E-01 | 2.13E-02 | 1.65E+02 | 1.35E-02 | 2.55E-03 |
| | Std | 1.19E+01 | 1.00E-02 | 1.84E-04 | 2.25E+00 | 1.74E-04 | 3.48E-05 |
| | Mean/Std | 116 | 66 | 116 | 73 | 78 | 73 |
| BL-VOS$_4$ | Nominal | 7.54E+03 | 3.19E+00 | 1.16E-01 | 1.65E+03 | 1.29E-01 | 2.54E-02 |
| | Mean | 7.60E+03 | 3.33E+00 | 1.22E-01 | 1.60E+03 | 1.18E-01 | 2.49E-02 |
| | Std | 6.21E+01 | 4.40E-02 | 1.00E-03 | 1.87E+01 | 1.23E-03 | 2.88E-04 |
| | Mean/Std | 122 | 76 | 122 | 86 | 96 | 86 |

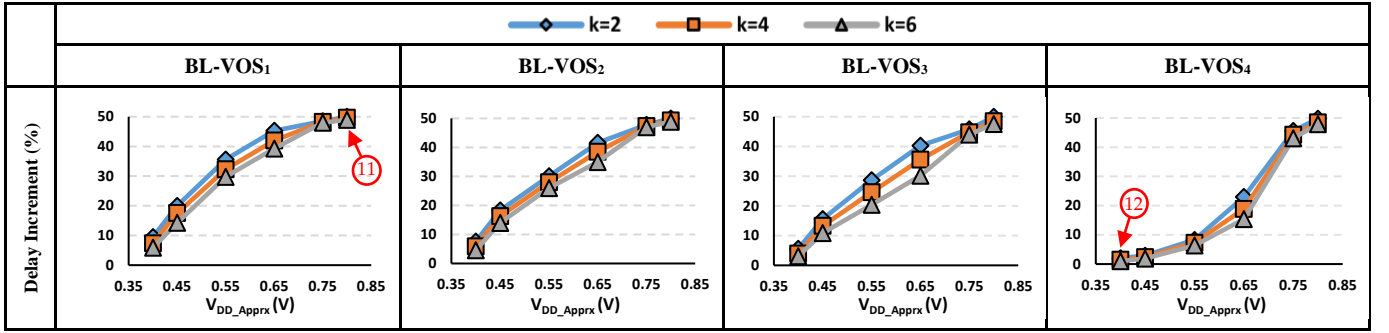

Fig. 13.  Variations of delay induced by the aging effects on the different structures of the BL-VOS multiplier for the different values of k and $V_{DD\_Apprx}$ after ten years of circuit operation.

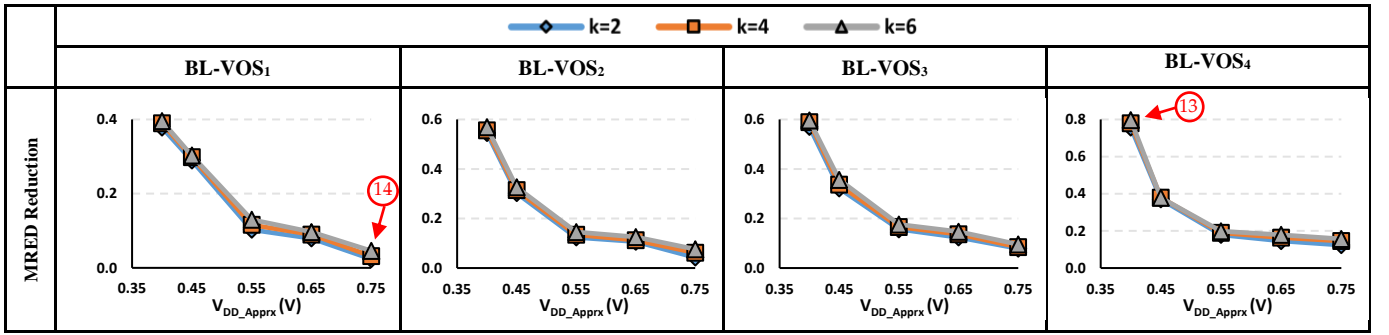

Fig. 14.  Variations of MRED induced by the delay effects on the different structures of the BL-VOS multiplier for the different values of k and $V_{DD\_Apprx}$ after ten years of circuit operation.

## 6. CONCLUSION

In this article, we presented a generic VOS-based accuracy configurable multiplier with improved energy consumption and lifetime. Thanks to applying VOS at the coarse level, our proposed BL-VOS multipliers achieved significantly better trade-offs in energy consumption and output accuracy compared to the state-of-the-art VOS-based multipliers. Also, the different structures of the proposed BL-VOS multiplier result in, on average, 57% lower energy consumption compared to the exact multiplier. Process variation and aging phenomenon effects on the proposed multipliers were also studied. Finally, we evaluated the various proposed BL-VOS multipliers in the realization of image processing applications. Based on the results, our proposed multipliers have shown up to 80% lower energy consumption, for only 4% output quality degradation compared to the exact multiplier.

TABLE X   MSSIM AND ENERGY REDUCTION OF THE DIFFERENT STRUCTURES OF THE BL-VOS-BASED APPROXIMATE MULTIPLIERS WITH k = 4 AND FOR TWO DIFFERENT APPROXIMATE VOLTAGE LEVELS

| | MSSIM | | | | | | | | | | | | | | | |
|---|---|---|---|---|---|---|---|---|---|---|---|---|---|---|---|---|
| Application | Sharpening | | | | | | | | Smoothing | | | | | | | |
| VOS Voltage Level | 0.4V | | | | 0.75V | | | | 0.4V | | | | 0.75V | | | |
| Structure | $BL\text{-}VOS_1$ | $BL\text{-}VOS_2$ | $BL\text{-}VOS_3$ | $BL\text{-}VOS_4$ | $BL\text{-}VOS_1$ | $BL\text{-}VOS_2$ | $BL\text{-}VOS_3$ | $BL\text{-}VOS_4$ | $BL\text{-}VOS_1$ | $BL\text{-}VOS_2$ | $BL\text{-}VOS_3$ | $BL\text{-}VOS_4$ | $BL\text{-}VOS_1$ | $BL\text{-}VOS_2$ | $BL\text{-}VOS_3$ | $BL\text{-}VOS_4$ |
| Female | 0.99 | 0.92 | 0.97 | 0.82 | 1 | 1 | 1 | 1 | 0.98 | 0.84 | 0.95 | 0.98 | 1 | 1 | 1 | 1 |
| House | 1 | 0.97 | 0.97 | 0.92 | 1 | 1 | 1 | 1 | 0.99 | 0.86 | 0.99 | 0.99 | 1 | 1 | 1 | 1 |
| Jelly beans | 0.98 | 0.92 | 0.94 | 0.88 | 1 | 1 | 1 | 1 | 0.99 | 0.85 | 1 | 0.98 | 1 | 1 | 1 | 1 |
| Baboon | 1 | 0.98 | 0.99 | 0.93 | 1 | 1 | 1 | 1 | 0.99 | 0.83 | 0.98 | 0.98 | 1 | 1 | 1 | 1 |
| Peppers | 1 | 0.99 | 0.99 | 0.99 | 1 | 1 | 1 | 1 | 1 | 0.94 | 1 | 1 | 1 | 1 | 1 | 1 |
| Energy Reduction (%) | 33.6 | 18.3 | 45.3 | 79.9 | 12.6 | 3.9 | 4.6 | 3.1 | 33.6 | 18.3 | 45.3 | 79.9 | 12.6 | 3.9 | 4.6 | 3.1 |

| | $k = 4$, $V_{DD} = 0.8V$ | $k = 4$, $V_{DD\_Apprx} = 0.55V$ | | | |
|---|---|---|---|---|---|
| Structure | Exact | $BL\text{-}VOS_1$ | $BL\text{-}VOS_2$ | $BL\text{-}VOS_3$ | $BL\text{-}VOS_4$ |
| Sharpening | 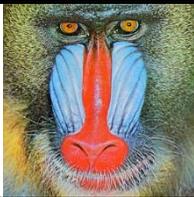 | 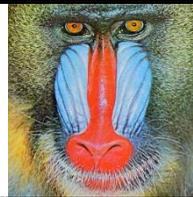 | 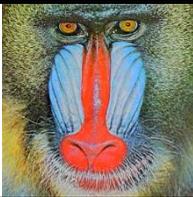 | 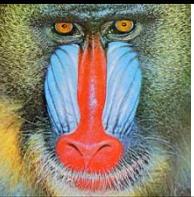 | 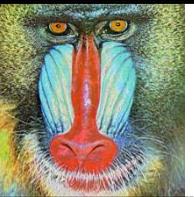 |
| MSSIM | 1 | 1 | 0.99 | 1 | 0.93 |
| Smoothing | 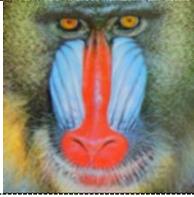 | 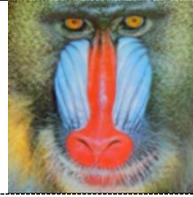 | 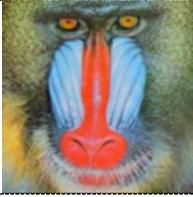 | 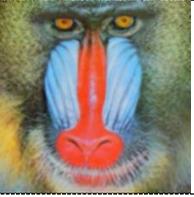 | 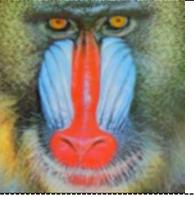 |
| MSSIM | 1 | 1 | 1 | 0.99 | 0.99 |

Fig. 15.  Output images of the sharpening and smoothing image processing applications for the "Baboon" image benchmark, for the exact and the different structures of BL-VOS multipliers with k = 4 and the voltage level of 0.55V.

## REFERENCES

[1] P. Kulkarni, P. Gupta and M. Ercegovac, "Trading accuracy for power with an underdesigned multiplier architecture," in *Proc. 24th Intl. Conf. on VLSI Des.*, Jan. 2011, pp. 346–351.

[2] O. Akbari, M. Kamal, A. Afzali-Kusha, and M. Pedram, "Dual-quality 4:2 compressors for utilizing in dynamic accuracy configurable multipliers," *IEEE Trans. Very Large Scale Integr. (VLSI) Syst.*, vol. 25, no. 4, pp. 1352–1361, Apr. 2017.

[3] H. Afzali-Kusha, M. Vaeztourshizi, M. Kamal and M. Pedram, "Design exploration of energy-efficient accuracy-configurable dadda multipliers with improved lifetime based on voltage overscaling," *IEEE Trans. Very Large Scale Integr. (VLSI) Syst.*, vol. 28, no. 5, pp. 1207-1220, May 2020.

[4] D. May and W. Stechele, "Voltage over-scaling in sequential circuits for approximate computing," in *Proc. 11th Int. Conf. Des. Technol. Integr. Syst. Nanosc. Era (DTIS)*, 2016, pp. 1-6.

[5] P. Chowdhury and B. C. Schafer, "Leveraging automatic high-level synthesis resource sharing to maximize dynamical voltage overscaling with error control," *ACM TOADES*, vol. 37, no. 4, Aug. 2021.

[6] A. Gupta, S. Mandavalli, V. J. Mooney, K.-V. Ling, A. Basu, H. Johan, and B. Tandianus, "Low power probabilistic floating-point multiplier design," in *Proc. IEEE ISVLSI*, Jul. 2011, pp. 182–187.

[7] M. Lau, K. V. Ling, and Y. C. Chu, "Energy-aware probabilistic multiplier: Design and analysis," in *Proc. CASES*, Oct. 2009, pp. 281–290.

[8] V. Gupta, D. Mohapatra, A. Raghunathan and K. Roy, "Low-power digital signal processing using approximate adders," *IEEE Trans. Comput.-Aided Des. Integr. Circuits Syst.*, vol. 32, no. 1, pp. 124-137, Jan. 2013.

[9] O. Akbari, M. Kamal, A. Afzali-Kusha, M. Pedram and M. Shafique, "X-CGRA: An energy-efficient approximate coarse-grained reconfigurable architecture," *IEEE Trans. Comput.-Aided Des. Integr. Circuits Syst.*, vol. 39, no. 10, pp. 2558-2571, Oct. 2020.

[10] W. Liu, F. Lombardi and M. Shulte, "A retrospective and prospective view of approximate computing [point of view]," in *Proc. IEEE*, vol. 108, no. 3, pp. 394-399, Mar. 2020.

[11] H. R. Mahdiani, A. Ahmadi, S. M. Fakhraie, and C. Lucas, "Bio-inspired imprecise computational blocks for efficient VLSI implementation of soft-computing applications," *IEEE Trans. Circuits Syst. I, Reg. Papers*, vol. 57, no. 4, pp. 850–862, Apr. 2010.

[12] V. Mrazek, R. Hrbacek, Z. Vasicek, and L. Sekanina, "EvoApprox8b: Library of approximate adders and multipliers for circuit design and benchmarking of approximation methods," in *Proc. Design, Autom. Test Eur. Conf. Exhib. (DATE)*, Mar. 2017, pp. 258–261.

[13] G. Zervakis, K. Tsoumanis, S. Xydis, D. Soudris, and K. Pekmestzi, "Design-efficient approximate multiplication circuits through partial product perforation," *IEEE Trans. Very Large Scale Integr. (VLSI) Syst.*, vol. 24, no. 10, pp. 3105–3117, Oct. 2016.

[14] D. Esposito, A. G. M. Strollo, E. Napoli, D. De Caro, and N. Petra, "Approximate multipliers based on new approximate compressors," *IEEE Trans. Cir. Syst. I, Reg. Papers*, vol. 65, no. 12, pp. 4169-4182, Dec. 2018.

[15] S. Narayanamoorthy, H. A. Moghaddam, Z. Liu, T. Park and N. S. Kim, "Energy-efficient approximate multiplication for digital signal processing and classification applications," *IEEE Trans. Very Large Scale Integr. (VLSI) Syst.*, vol. 23, no. 6, pp. 1180-1184, June 2015.

[16] A. Momeni, J. Han, P. Montuschi, and F. Lombardi, "Design and analysis of approximate compressors for multiplication," *IEEE Trans. Comput.*, vol. 64, no. 4, pp. 984–994, Apr. 2015.

[17] T. Yang, T. Ukezono, and T. Sato, "A low-power high-speed accuracy-controllable approximate multiplier design," in *Proc. 23rd Asia South Pacific Design Autom. Conf. (ASP-DAC)*, Jan. 2018, pp. 605–610.

[18] H. Baba, T. Yang, M. Inoue, K. Tajima, T. Ukezono, and T. Sato, "A low-power and small-area multiplier for accuracy-scalable approximate computing," in *Proc. IEEE ISVLSI*, Jul. 2018, pp. 569–574.

[19] B. Bailey. (2021). Chip Aging Becomes Design Problem. Accessed: Sep. 2021. [Online]. Available: https://semiengineering.com/chip-aging-becomes-design-problem/

[20] J. W. McPherson, Reliability Physics and Engineering Time-to-Failure Modeling, 2nd ed. Boston, MA, USA: Springer, 2013.

[21] E. Cai, D. Stamoulis, and D. Marculescu, "Exploring aging deceleration in FinFET-based multi-core systems," in *Proc. 35th Int. Conf. Comput. Aided Design (ICCAD)*, Nov. 2016, pp. 1–8.

[22] W. J. Townsend, E. E. Swartzlander and J. A. Abraham, "A comparison of Dadda and Wallace multiplier delays," in *Proc. SPIE Symp. Adv. Signal Process. Algorithms Archit. Implem.*, vol. 5205, Dec., 2003, pp. 552-560.

[23] W. Liu, L. Qian, C. Wang, H. Jiang, J. Han, and F. Lombardi, "Design of approximate Radix-4 booth multipliers for error-tolerant computing," *IEEE Trans. Comput.*, vol. 66, no. 8, pp. 1435–1441, Aug. 2017.

[24] H. Kaul, M. Anders, S. Hsu, A. Agarwal, R. Krishnamurthy, and S. Borkar, "Near-threshold voltage (NTV) design: Opportunities and challenges," in *Proc. IEEE DAC*, 2012, pp. 1149–1154.

[25] S. Kabirpour and M. Jalali, "A power-delay and area efficient voltage level shifter based on a reflected-output wilson current mirror level shifter," *IEEE Trans. Circuits Syst. II Exp. Briefs*, vol. 67, no. 2, pp. 250-254, Feb. 2020.

[26] A. Shapiro and E. G. Friedman, "Power efficient level shifter for 16 nm FinFET near threshold circuits," *IEEE Trans. Very Large Scale Integr. (VLSI) Syst.*, vol. 24, no. 2, pp. 774–778, Feb. 2016.

[27] A. Shafaei, Y. Wang, L. Chen, S. Chen, and M. Pedram, "Maximizing the performance of NoC-based MPSoCs under total power and power density constraints," in *Proc. 17th Int. Symp. Qual. Electron. Design (ISQED)*, Mar. 2016, pp. 49–56.

[28] H. Afzali-Kusha, O. Akbari, M. Kamal, and M. Pedram, "Energy consumption and lifetime improvement of coarse-grained reconfigurable architectures targeting low-power error-tolerant applications," in *Proc. Great Lakes Symp. VLSI (GLSVLSI)*, 2018, pp. 431–434.

[29] M. Martins, J. M. Matos, R. P. Ribas, A. Reis, G. Schlinker, L. Rech, and J. Michelsen, "Open cell library in 15nm FreePDK technology," in *Proc. Symp. Int. Symp. Phys. Design (ISPD)*, 2015, pp. 171–178.

[30] S. Carver, A. Mathur, L. Sharma, P. Subbarao, S. Urish and Q. Wang, "Low-power design using the Si2 common power format," *IEEE Design and Test of Computers*, vol. 29, no. 2, pp. 62-70, Apr. 2012.

[31] P. Pandey, P. Basu, K. Chakraborty, and S. Roy, "GreenTPU: Improving timing error resilience of a near-threshold tensor processing unit," in *Proc. 56th Annu. Design Autom. Conf. (DAC)*, 2019, pp. 1–6.

[32] M. Ansari, H. Afzali-Kusha, B. Ebrahimi, Z. Navabi, A. Afzali-Kusha, and M. Pedram, "A near-threshold 7T SRAM cell with high write and read margins and low write time for sub-20 nm FinFET technologies," *Integration*, vol. 50, pp. 91–106, Jun. 2015.

[33] H. R. Myler and A. R. Weeks, The Pocket Handbook of Image Processing Algorithms. Englewood Cliffs, NJ, USA: Prentice-Hall, 2009.

[34] Sipi.usc.edu. (2021). SIPI Image Database. [Online]. Available: http://sipi.usc.edu/database/

[35] J. Rabaey, Low Power Design Essentials. New York, NY 10013, USA: Springer, 2009.

[36] S. Rehman, W. El-Harouni, M. Shafique, A. Kumar and J. Henkel, "Architectural-space exploration of approximate multipliers," in *Proc. IEEE International Conference on Computer-Aided Design (ICCAD)*, pp. 1-8, 2016.

[37] V. Mrazek, M. A. Hanif, Z. Vasicek, L. Sekanina and M. Shafique, "autoAx: An Automatic Design Space Exploration and Circuit Building Methodology utilizing Libraries of Approximate Components," in *Proc. 56th IEEE Design Automation Conference (DAC)*, pp. 1-6, 2019.

[38] O. Akbari, M. Kamal, A. Afzali-Kusha, M. Pedram and M. Shafique, "Toward Approximate Computing for Coarse-Grained Reconfigurable Architectures," in *IEEE Micro*, vol. 38, no. 6, pp. 63-72, 1 Nov.-Dec. 2018.

[39] I. Koren, Computer arithmetic algorithms. Natick, Ma: A K Peters, 2002.

[40] M. Traiola, A. Virazel, P. Girard, M. Barbareschi and A. Bosio, "A Survey of Testing Techniques for Approximate Integrated Circuits," in *Proc. IEEE*, vol. 108, no. 12, pp. 2178-2194, Dec. 2020.

[41] M. Traiola, A. Virazel, P. Girard, M. Barbareschi and A. Bosio, "Testing approximate digital circuits: Challenges and opportunities," in *Proc. IEEE 19th Latin-American Test Symposium (LATS)*, 2018, pp. 1-6.

[42] S. Hong, S. Kim, and W. E. Stark, "Low-power Application-specific Parallel Array Multiplier Design for DSP Applications," *VLSI Design*, vol. 14, no. 3, pp. 287–298, Jan. 2002.

[43] F. A. Shah, H. Jamal and M. A. Khan, "Reconfigurable Low Power FIR Filter based on Partitioned Multipliers," in *Proc. International Conference on Microelectronics*, 2006, pp. 87-90.

[44] A. Arora, Z. Wei and L. K. John, "Hamamu: Specializing FPGAs for ML Applications by Adding Hard Matrix Multiplier Blocks," in *Proc. IEEE 31st International Conference on Application-specific Systems, Architectures and Processors (ASAP)*, 2020, pp. 53-60.

[45] S. Hauck and André Dehon, Reconfigurable computing: the theory and practice of FPGA-based computation. Amsterdam; Boston: Morgan Kaufmann, 2008.

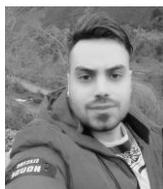
**Ali Akbar Bahoo** received his M.Sc. from the University of Eyvanekey, Iran, in 2021. His research interests include approximate computing, deep neural networks (DNNs), electronic design automation, low-power design, hardware/software co-design, and computer architecture.

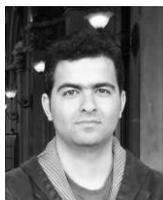
**Omid Akbari** received his Ph.D. in electrical engineering, digital systems sub-discipline from the University of Tehran, Iran, in 2018,. He is currently an assistant professor with the School of Electrical and Computer Engineering at the Tarbiat Modares University, Iran. His current research interests include energy-efficient computing, reliability, and machine learning.

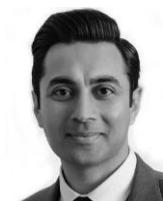
**Muhammad Shafique** (SM'16) received his Ph.D. in computer science from Karlsruhe Institute of Technology, Germany, in 2011. He is currently with the Division of Engineering at New York University (NYU) Abu Dhabi. His research interests are in system-level design AI/Machine Learning, brain-inspired computing, wearables, autonomous systems, and Smart CPS.